\pdfoutput=1
\RequirePackage{ifpdf}
\ifpdf % We are running pdfTeX in pdf mode
\documentclass[pdftex]{sigma}
\else
\documentclass{sigma}
\fi

\numberwithin{equation}{section}
\newtheorem{Theorem}{Theorem}[section]

\usepackage{tikz}
\usetikzlibrary{calc}

\begin{document}

\allowdisplaybreaks

\newcommand{\arXivNumber}{1312.1440}

\renewcommand{\thefootnote}{}

\renewcommand{\PaperNumber}{041}

\FirstPageHeading

\ShortArticleName{A Variational Principle for Discrete Integrable Systems}

\ArticleName{A Variational Principle for Discrete Integrable\\ Systems\footnote{This paper is a~contribution to the Special Issue on Recent Advances in Quantum Integrable Systems. The full collection is available at \href{http://www.emis.de/journals/SIGMA/RAQIS2016.html}{http://www.emis.de/journals/SIGMA/RAQIS2016.html}}}

\Author{Sarah B.~LOBB~$^\dag$ and Frank W.~NIJHOFF~$^\ddag$}

\AuthorNameForHeading{S.B.~Lobb and F.W.~Nijhoff}

\Address{$^\dag$~NSW Department of Education, Sydney NSW 2000, Australia}
\EmailD{\href{mailto:sarah.lobb.education@gmail.com}{sarah.lobb.education@gmail.com}}

\Address{$^\ddag$~School of Mathematics, University of Leeds, Leeds LS2 9JT, UK}
\EmailD{\href{mailto:F.W.Nijhoff@leeds.ac.uk}{F.W.Nijhoff@leeds.ac.uk}}

\ArticleDates{Received April 01, 2017, in final form April 26, 2018; Published online May 03, 2018}

\Abstract{For integrable systems in the sense of multidimensional consistency (MDC) we can consider the Lagrangian as a form, which is closed on solutions of the equations of motion. For 2-dimensional systems, described by partial difference equations with two independent variables, MDC allows us to define an action on arbitrary 2-dimensional surfaces embedded in a higher dimensional space of independent variables, where the action is not only a functional of the field variables but also  the choice of surface. It is then natural to propose that the system should be derived from a variational principle which includes not only variations with respect to the dependent variables, but also with respect to variations of the surface in the space of independent variables. Here we derive the resulting system of generalized Euler--Lagrange equations arising from that principle. We treat the case where the equations are 2 dimensional (but which due to MDC can be consistently embedded in higher-dimensional space), and show that they can be integrated to yield relations of quadrilateral type. We also derive the extended set of Euler--Lagrange equations for 3-dimensional systems, i.e., those for equations with 3 independent variables. The emerging point of view from this study is that the variational principle can be considered as the set of equations not only encoding the equations of motion but as the defining equations for the Lagrangians themselves.}


\Classification{35Q51; 37K60; 39A14; 49N99}

\newcommand{\drawTriangle}[3]
{
 \filldraw[fill=green!40!,draw=black,very thick] (#1) -- (#2) -- (#3) -- (#1);
}
\newcommand{\drawShadySquare}[4]
{
 \filldraw[fill=green!20!,draw=black] (#1) -- (#2) -- (#3) -- (#4) -- (#1);
}
\newcommand{\drawSquare}[4]
{
 \filldraw[fill=gray!80!,draw=black,very thick] (#1) -- (#2) -- (#3) -- (#4) -- (#1);
}
\newcommand{\drawGridLine}[2]
{
 \draw[gray,very thin] (#1) -- (#2);
}
\newcommand{\drawDot}[1]
{
 \shade[ball color=black] (#1) circle (0.08);
}
\newcommand{\drawDiamond}[1]
{
 \shade[ball color=black] (#1) circle (0.08);
}
\newcommand{\drawGridLeft}
{
 \filldraw[fill=gray!20,fill opacity=0.5] (0,1) rectangle (0.6,0);
 \drawGridLine{0.3,1}{0.3,0}
 \drawGridLine{0,0.5}{0.6,0.5}
 \drawGridLine{0.8,0.375}{0.8,-0.625}
 \drawGridLine{0.5,-0.125}{1.1,-0.125}
}
\newcommand{\drawGridBottom}
{
 \filldraw[fill=gray!50,fill opacity=0.5] (0,0) rectangle (1.25,0.6);
 \filldraw[fill=gray!50,fill opacity=0.5] (-0.5,0.4) rectangle (0.75,1);
 \drawGridLine{0.625,0}{0.625,0.6}
 \drawGridLine{0,0.3}{1.25,0.3}
 \drawGridLine{0.125,0.4}{0.125,1}
 \drawGridLine{-0.5,0.7}{0.75,0.7}
}
\newcommand{\drawGridBack}
{
 \filldraw[fill=gray!20,fill opacity=0.5] (0.6,0.75) rectangle (1.6,1.75);
 \drawGridLine{1.1,0.75}{1.1,1.75}
 \drawGridLine{0.6,1.25}{1.6,1.25}
 \drawGridLine{0.8,0.375}{0.8,1.375}
 \drawGridLine{0.3,0.875}{1.3,0.875}
}
\newcommand{\drawGridRight}
{
 \filldraw[fill=gray!20,fill opacity=0.5] (1,-0.25) rectangle (1.6,-1.25);
 \drawGridLine{1.3,-0.25}{1.3,-1.25}
 \drawGridLine{1,-0.75}{1.6,-0.75}
}
\newcommand{\drawGridTop}
{
 \filldraw[fill=gray!50,fill opacity=0.5] (-1,0.8) rectangle (0.25,1.4);
 \drawGridLine{-0.375,0.8}{-0.375,1.4}
 \drawGridLine{-1,1.1}{0.25,1.1}
}
\newcommand{\drawGridFront}
{
 \filldraw[fill=gray!20,fill opacity=0.5] (0,0) rectangle (1,1);
 \drawGridLine{0.5,0}{0.5,1}
 \drawGridLine{0,0.5}{1,0.5}
}

\renewcommand{\thefootnote}{\arabic{footnote}}
\setcounter{footnote}{0}

\section{Introduction}

It is an idea going back to the early 1700s that a physical system should minimize some quantity representing the work done by that system, i.e., in modern language, that the equations of motion of the corresponding dynamical system should arise as critical points of some action functional. This action is a quantity depending in principle on both the dependent and independent variables, and finding a minimum, or more generally a critical point, specifies a path expressed in terms of the dependent variable as a function of the independent (time-)variable. The condition, obtained by variational calculus, that specifies this path is the Euler--Lagrange (EL) equation.

A discrete calculus of variations, i.e., where the independent variable takes on discrete or integer values, was first developed outside the scope of integrable systems in the 1970s by Cadzow~\cite{Cad1970}, Logan~\cite{Log1973} in the context of control theory, and later for dynamical systems by Maeda~\cite{Mae1980,Mae1981,Mae1982}. Cadzow's original motivation was the use of the digital computer in modern systems and the solution of control problems, and it became clear that the formulation of a~discrete calculus of variations was important for numerical methods, in optimization and engineering problems. In the discrete realm instead of the action being an integral of a Lagrangian function over the time-variable, it is here a sum over the independent discrete variable.

A modern notion of integrability for higher-dimensional systems, i.e., systems described by partial differential equations (PDEs) or partial difference equations (P$\Delta$Es) is that of multidimensional consistency, cf., e.g.,~\cite{HietJoshNij16}. In the case of multidimensionally consistent systems, we are able to embed the system in a higher-dimensional space, with compatible systems of equations defined in each subspace of dimensionality corresponding to the number of independent variables of the given equation. In general, we may have an infinite number of compatible systems in an infinite number of dimensions, and we do not need to restrict ourselves to any particular subspace. Furthermore, we could even have a~system following a path through an arbitrary number of dimensions. In that case we have to consider not only the path taken by the dependent variable(s) with respect to the independent variable(s), but also the path through this space of independent variables. Then it is natural to ask that the action be critical with respect to changes in the path in the space of independent variables.

This postulate was first put forward by the authors in \cite{LobNij2009}, initially for 2-dimensional systems, both discrete and continuous. Requiring the action functional to be invariant under small changes in the path (which for 2-dimensional systems is actually a surface) through the space of independent variables leads to a condition on the Lagrangian, namely a closure relation, which was shown to be satisfied for many examples of multidimensionally consistent systems~\cite{BobSur2010,BolSur2012,LobNij2009,LobNij2010,LobNijQui2009,XenNijLob2011}. This then serves as an answer to the question of how to encode an entire multidimensionally consistent system within a single variational set-up (in contrast to the conventional variational situation that the EL equation only provides one single equation of motion for each component of the field variables, and not a compatible system of equations for one and the same field variable).

A technical issue with the usual variational principle is that it is often not possible to obtain the desired system of equations as Euler--Lagrange equations, but only a weaker (integrated or derivative) form of those equations. This can be seen in the continuous realm for instance in the case of the potential Korteweg--de Vries (pKdV) equation, where the Euler--Lagrange equation gives only a derivative form of the pKdV, which is the actual (non-potential) KdV equation, even when we vary with respect to the pKdV variable. In the discrete realm a parallel situation occurs in the case of quadrilateral lattice equations (quad equations), such as those in the Adler--Bobenko--Suris (ABS) classification \cite{AdlBobSur2003}, where the EL equations only yield a weak form (a~difference consequence) of the quad equation, which can be considered as its discrete `derivative'. Thus, one often does not obtain the quad equation itself directly as an Euler--Lagrange equation on a fixed surface.

We show in this paper that the variational principle of \cite{LobNij2009}, which considers variations of the surface in the space of independent variables as well as variations of the dependent variables, provides a system of generalized Euler--Lagrange (EL) equations, which comprises conditions on both the Lagrangian as well as the solutions of the equations of motion. In the 2-dimensional discrete case, starting from this system of EL equations with 3-point Lagrangians depending on a scalar field variable~$u$, we derive under the assumption that the equations hold in any number of lattice directions, by a discrete integration procedure quadrilateral type lattice equations opossessing the MDC property. These quad equations may be interpreted as conservation laws for the EL equations, and may still depend on functions of the lattice sites but in a way that doesn't single out any one of the lattice directions (this is an essential freedom in MDC quad equations). However, the key point we wish to bring across is that the variational principle should be considered not only as a mechanism to encode the equations of motion by means of specifying a given Lagrangian, but rather as a principle defining a system of partial differential equations for the Lagrangian components of a ``Lagrangian multiform''. In principle, one would like to establish the general solution for the Lagrangian of this system of generalised EL equations, but in the present paper we shall consider only the restrictive class of solutions of 3-point Lagrangians, and show that under additional assumptions (e.g., that Lagrangians are quadratic) they are determined almost uniquely. Once the Lagrangians are established, the resulting quad equation necessarily possesses the MDC property. Thus, the integrability (in the sense of MDC) emerges in an organic way from the extended variational principle, and it is the latter perspective that forms the main departure of our new variational principle from existing conventional variational theories.

The case of 1-dimensional systems this phenomenon was first examined in \cite{Yoo2011, YooLobNij2010,YooNij2013}, where the new varational point of view was first proposed, followed subsequently by further work in \cite{BolPetSur2013a,BolPetSur2015a, Sur2012} from a more Hamiltonian perspective. The present study on 2-dimensional MDC systems parallels related work, albeit from quite a different perspective, of~\cite{BolPetSur2013b} which was conducted independently and simultaneously, cf.~\cite{Nij2013}.

This paper is mainly concerned with discrete systems, but naturally the analogous continuous situation, where the independent variables of the equations are continuous, was developed as well, cf.~\cite{LobNij2009,XenNijLob2011}, while in~\cite{SurVerm2015} the variational system for hierarchies of integrable PDEs was set up. Some of the results of the present paper were also summarised in~\cite[Chapter~12]{HietJoshNij16} but without any of the computational details. The organisation of the paper is as follows: In Section~\ref{section_2d} we examine the variational principle for 2-dimensional discrete systems: defining the action, deriving the set of extended EL equations for the basic configurations of the surface, and performing the analysis of the EL equations to yield the quad equations. We present the examples of the H1 and H3 quad equations, cf.~\cite{AdlBobSur2003}, as illustrations, and furthermore show that in the case of quadratic Lagrangians the analysis yields a unique parameter class of Lagrangians corresponding to a linear quad equation possessing the MDC property. In Section~\ref{section_3d} we give the defining set of EL equations for 3-dimensional discrete systems, and show that these are compatible with the bilinear discrete Kadomtsev--Petviashvili (KP) equation. Section \ref{section_conclusion} provides some further discussion and perspectives.

\section{Discrete 2-dimensional systems}\label{section_2d}
In this section we will set up the variational multiform approach for the case of 2-dimensional discrete systems, i.e., for systems of P$\Delta$Es with two independent discrete variables corresponding to field configurations around an elementary quadrilateral. For a large class of equations defined on a quadrilateral, namely those in the ABS list \cite{AdlBobSur2003}, we have Lagrangians involving 3 points of the quadrilateral. We will define the action on arbitrary quad-surfaces for such systems, when the lattice on which the field variables are defined, is embedded in a lattice of arbitrary higher (i.e., $>2$) dimensionality, or in a higher-dimensional ambient space. Instead of quad-equations on a~regular 2-dimensional lattice we consider equations of the motion, which, while still involving only two independent variables labelling the vertices of each quadrilateral of an arbitrary surface configuration, may be associated with different lattice directions at different quadrilaterals. The simultaneous consistency of these equations for different choices of the surface is expressed by the MDC property, allowing quad equations in different directions to be imposed simultaneously on one and the same field variable at each vertex of the surface.

\subsection{Defining the action}
\begin{figure}[h]\centering
\begin{tikzpicture}[scale=2]
\begin{scope}[every node/.append style={xslant=0,yslant=-0.25},xslant=0,yslant=-0.25]
 \filldraw[fill=gray!20,fill opacity=0.5] (1.3,0.875) rectangle (0.8,1.375);
 \draw[thick,->] (0.8,0.875) -- (1.3,0.875); \node at (1.2,0.75) {$\boldsymbol{e}_{i}$};
 \draw[thick,->] (0.8,0.875) -- (0.8,1.375); \node at (0.7,1.25) {$\boldsymbol{e}_{j}$};
\end{scope}
 \drawDiamond{0.8,0.675}
 \node[white] at (0.8,0.675) {$\boldsymbol{n}$};
\end{tikzpicture}
\vspace*{-5mm}

\caption{Elementary oriented plaquette.}\label{plaquette}
\end{figure}
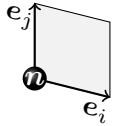

To set up the action functional we consider the sum of Lagrangian contributions from all elementary plaquettes in a given quad-surface. To make this precise, we consider the quad-surface~$\sigma$ to be a connected configuration of elementary plaquettes $\sigma_{ij}(\boldsymbol{n})$, where $\sigma_{ij}(\boldsymbol{n})$ is specified by the position $\boldsymbol{n}$ of one of the vertices in the lattice relative to a given reference frame of arbitrary dimension\footnote{Although not necessary, we can think of a quadrilateral surface embedded in $\mathbb{Z}^d$ where $d=|I|$, and where the coordinates of the lattice vertices are given by $\boldsymbol{n}=(n_1,n_2, \dots,n_d)$.}, and configurations $(\boldsymbol{n},\boldsymbol{n}+\boldsymbol{e}_{i},\boldsymbol{n}+\boldsymbol{e}_{j})$ defining a~quadrilateral with elementary lattice directions given by base vectors $\boldsymbol{e}_{i}$, $\boldsymbol{e}_{j}$ ($i$, $j$ denoting labels running over a chosen index set~$I$ of cardinality $>2$), as in Fig.~\ref{plaquette}. The surface can be closed, or have a fixed boundary.

We associate with each of the lattice directions given by the base vectors $\boldsymbol{e}_i$, $i\in I$, a distinct parameter $\alpha_i$ taking values in some appropriate set of numbers. (In some cases these parameters can take as values points on an algebraic curve, and they can vary along the lattice, but only along the direction in the associated lattice direction, i.e., $\alpha_i=\alpha_i(n_i)$. %

Since the 3-point Lagrangians depend on two directions in the lattice, we can associate with it, at each point of a multidimensional lattice, an
oriented plaquette $\sigma_{ij}(\boldsymbol{n})$. Thus, we can think of these Lagrangians as defining a discrete 2-form $\mathcal{L}_{ij}(\boldsymbol{n})$ whose evaluation on that plaquette is given by the Lagrangian function as follows
\begin{gather}
\mathcal{L}_{ij}(\boldsymbol{n}) = L(u(\boldsymbol{n}),u(\boldsymbol{n}+\boldsymbol{e}_{i}),u(\boldsymbol{n}+\boldsymbol{e}_{j});\alpha_{i},\alpha_{j}). \label{Lagrangian}
\end{gather}
Here, the $u(\boldsymbol{n})$ are the (variational) field variables depending on the lattice sites $\boldsymbol{n}$, while the $\alpha_i$ are lattice parameters associated with the direction labelled by the index $i$ in a lattice of arbitrary dimension. The Lagrangians, as the ones given in \cite{LobNij2009}, are postulated to be anti-symmetric (skew symmetric) with respect to the interchange of lattice directions $i$, $j$, $\mathcal{L}_{ij}=-\mathcal{L}_{ji}$, and so the orientation is well-defined. Consequently the action $S$ is also well-defined by
 \begin{gather}
 S[u(\boldsymbol{n});\sigma] = \sum_{\sigma_{ij}(\boldsymbol{n})\in\sigma}{\mathcal{L}_{ij}(\boldsymbol{n})},\label{action}
 \end{gather}
where in performing the lattice sum the orientation of the plaquettes is taken into account.

\subsection{The Euler--Lagrange equations}

To derive the set of Euler--Lagrange equations stemming from the action \eqref{action}, we look at what happens at a particular point $\boldsymbol{n}$ in the lattice. For ease of notation we will suppress the dependence on $\boldsymbol{n}$, writing $u=u(\boldsymbol{n})$, and make use of shift operators $T_{i}$, writing $T_{i}u=u(\boldsymbol{n}+\boldsymbol{e}_{i})$, $T_{j}u=u(\boldsymbol{n}+\boldsymbol{e}_{j})$, $T_{i}^{-1}u=u(\boldsymbol{n}-\boldsymbol{e}_{i})$, etc.

The central postulate is that the system, defined by the equations for $u$, lies at a critical point of the action \eqref{action} viewed as a functional of both the field variable~$u$ and of the surface~$\sigma$. Since we are considering discrete surfaces here, the notion of infinitesimal variations of the independent variables does not quite make sense, and hence we consider only finite local variations of the discrete surface by flips of quadrilaterals over elementary cubes (or rhomboids). Thus, our postulate is that the action is independent of~$\sigma$ (while keeping any boundary $\partial\sigma$ of the surface fixed) on solutions of the system of equations for~$u$.

To set up the fundamental system of EL equations it suffices to consider a collection of elementary surfaces embedded in 3 dimensions, and compute variations with respect to $u$ on that surface. For an action which is the sum of 3-point Lagrangians $L(u,T_{i}u,T_{j}u;\alpha_{i},\alpha_{j})$, there are various possible configurations involving the arbitrary point $u$. The most obvious elementary (i.e., minimal containing a central point) configuration is the flat 2-dimensional one comprising three quads meeting at the central point, as shown in Fig.~\ref{2d_config}:

\begin{figure}[h]\centering
%%%%%%%%%%%%%%%%%%
%%%	FLAT 2D SPACE	%%%
%%%%%%%%%%%%%%%%%%
\begin{tikzpicture}[every node/.style={minimum size=1cm},scale=2]
%%% back faces %%%
%% left face %%
\begin{scope}[every node/.append style={xslant=0,yslant=1},xslant=0,yslant=1]
 \drawGridLeft
\end{scope}
%% bottom face %%
\begin{scope}[every node/.append style={xslant=1,yslant=-0.2},xslant=1,yslant=-0.2]
 \drawGridBottom
\end{scope}
%% back face %%
 \begin{scope}[every node/.append style={xslant=0,yslant=-0.25},xslant=0,yslant=-0.25]
 \drawGridBack
 %%% Note that the central point in this grid is at 0.8,0.875 %%%
 %% Top left triangle %%
 \drawTriangle{0.8,0.875}{0.3,0.875}{0.3,1.375}
 %% Top right triangle %%
 \drawTriangle{1.3,0.875}{0.8,0.875}{0.8,1.375};
 %% Bottom right triangle %%
 \drawTriangle{1.3,0.375}{0.8,0.375}{0.8,0.875};
\end{scope}
%% circles %%
 \drawDiamond{0.8,0.675}
 \drawDot{0.3,0.775}
 \drawDot{0.8,0.175}
%%% front faces %%%
%% right face %%
\begin{scope}[every node/.append style={xslant=0,yslant=1},xslant=0,yslant=1]
 \drawGridRight
\end{scope}
%% top face %%
\begin{scope}[every node/.append style={xslant=1,yslant=-0.2},xslant=1,yslant=-0.2]
 \drawGridTop
\end{scope}
 %% front face %%
\begin{scope}[every node/.append style={xslant=0,yslant=-0.25},xslant=0,yslant=-0.25]
 \drawGridFront
\end{scope}
%% circles %%
 \drawDot{1.3,0.075}
 \drawDot{0.8,1.175}
 \drawDot{0.3,1.275}
 \drawDot{1.3,0.575}
\end{tikzpicture}\vspace{-2mm}
\caption{Usual configuration in 2 dimensions.}\label{2d_config}
\end{figure}
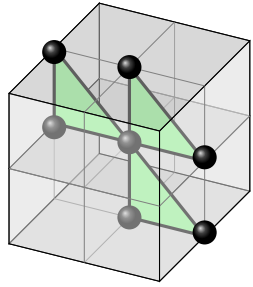

\begin{figure}[h]\centering
%%%%%%%%%%%%%%%%%%
%%%	PICTURE ONE		%%%
%%%%%%%%%%%%%%%%%%
\begin{tikzpicture}[every node/.style={minimum size=1cm},scale=2]
%%% back faces %%%
%% left face %%
\begin{scope}[every node/.append style={xslant=0,yslant=1},xslant=0,yslant=1]
 \drawGridLeft
\end{scope}
%% bottom face %%
\begin{scope}[every node/.append style={xslant=1,yslant=-0.2},xslant=1,yslant=-0.2]
 \drawGridBottom
\end{scope}
 %% back face %%
 \begin{scope}[every node/.append style={xslant=0,yslant=-0.25},xslant=0,yslant=-0.25]
 \drawGridBack
 %% ab-triangle %%
 \drawTriangle{1.3,0.875}{0.8,0.875}{0.8,1.375}
 %% bc-triangle %%
 \drawTriangle{0.8,0.875}{0.5,0.5}{0.8,1.375}
 %% ca-triangle %%
 \drawTriangle{0.8,0.875}{0.5,0.5}{1.3,0.875}
\end{scope}
%% circles %%
 \drawDiamond{0.8,0.675}
%%% front faces %%%
%% right face %%
\begin{scope}[every node/.append style={xslant=0,yslant=1},xslant=0,yslant=1]
 \drawGridRight
\end{scope}
%% top face %%
\begin{scope}[every node/.append style={xslant=1,yslant=-0.2},xslant=1,yslant=-0.2]
 \drawGridTop
\end{scope}
 %% front face %%
\begin{scope}[every node/.append style={xslant=0,yslant=-0.25},xslant=0,yslant=-0.25]
 \drawGridFront
\end{scope}
%% circles %%
 \drawDot{0.8,1.175}
 \drawDot{1.3,0.575}
 \drawDot{0.5,0.375}
\end{tikzpicture}
%%%%%%%%%%%%%%%%%%
%%%	PICTURE TWO		%%%
%%%%%%%%%%%%%%%%%%
\begin{tikzpicture}[every node/.style={minimum size=1cm},scale=2]
%%% back faces %%%
%% left face %%
\begin{scope}[every node/.append style={xslant=0,yslant=1},xslant=0,yslant=1]
 \drawGridLeft
\end{scope}
%% bottom face %%
\begin{scope}[every node/.append style={xslant=1,yslant=-0.2},xslant=1,yslant=-0.2]
 \drawGridBottom
\end{scope}
 %% back face %%
 \begin{scope}[every node/.append style={xslant=0,yslant=-0.25},xslant=0,yslant=-0.25]
 \drawGridBack
 %% ca-triangle %%
 \drawTriangle{0,0.5}{0.8,0.875}{0.3,0.875}
 %% ab-triangle %%
 \drawTriangle{0.8,0.875}{0.3,0.875}{0.3,1.375}
 %% bc-triangle %%
 \drawTriangle{0.8,1.375}{0.8,0.875}{0.5,0.5}
\end{scope}
%% circles %%
 \drawDiamond{0.8,0.675}
 \drawDot{0.3,0.775}
%%% front faces %%%
%% right face %%
\begin{scope}[every node/.append style={xslant=0,yslant=1},xslant=0,yslant=1]
 \drawGridRight
\end{scope}
%% top face %%
\begin{scope}[every node/.append style={xslant=1,yslant=-0.2},xslant=1,yslant=-0.2]
 \drawGridTop
\end{scope}
 %% front face %%
\begin{scope}[every node/.append style={xslant=0,yslant=-0.25},xslant=0,yslant=-0.25]
 \drawGridFront
\end{scope}
%% circles %%
 \drawDot{0.3,1.275}
 \drawDot{0.8,1.175}
 \drawDot{0.5,0.375}
 \drawDot{0,0.475}
\end{tikzpicture}
%%%%%%%%%%%%%%%%%%
%%%	PICTURE THREE	%%%
%%%%%%%%%%%%%%%%%%
\begin{tikzpicture}[every node/.style={minimum size=1cm},scale=2]
%%% back faces %%%
%% left face %%
\begin{scope}[every node/.append style={xslant=0,yslant=1},xslant=0,yslant=1]
 \drawGridLeft
\end{scope}
%% bottom face %%
\begin{scope}[every node/.append style={xslant=1,yslant=-0.2},xslant=1,yslant=-0.2]
 \drawGridBottom
\end{scope}
 %% back face %%
 \begin{scope}[every node/.append style={xslant=0,yslant=-0.25},xslant=0,yslant=-0.25]
 \drawGridBack
 %% ca-triangle %%
 \drawTriangle{0,0.5}{0.8,0.875}{0.3,0.875}
 %% No ab-triangle %%
 %% bc-triangle %%
 \drawTriangle{0.8,0.875}{0.8,0.375}{0.5,0}
\end{scope}
%% circles %%
 \drawDiamond{0.8,0.675}
 \drawDot{0.3,0.775}
 \drawDot{0.8,0.175}
%%% front faces %%%
%% right face %%
\begin{scope}[every node/.append style={xslant=0,yslant=1},xslant=0,yslant=1]
 \drawGridRight
\end{scope}
%% top face %%
\begin{scope}[every node/.append style={xslant=1,yslant=-0.2},xslant=1,yslant=-0.2]
 \drawGridTop
\end{scope}
 %% front face %%
\begin{scope}[every node/.append style={xslant=0,yslant=-0.25},xslant=0,yslant=-0.25]
 \drawGridFront
\end{scope}
%% circles %%
 \drawDot{0.5,-0.125}
 \drawDot{0,0.475}
\end{tikzpicture}\vspace{-2mm}
\caption{Elementary configurations in 3 dimensions.}\label{3d_configs}
\end{figure}
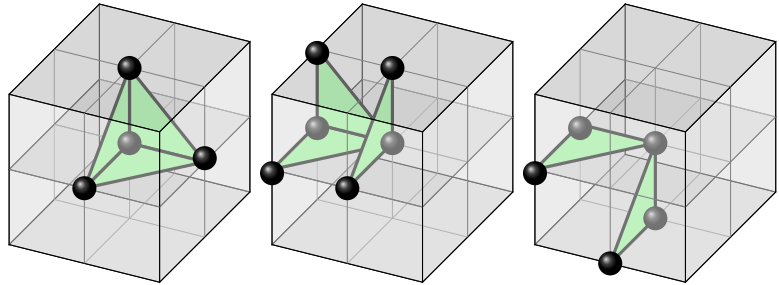

The corresponding Euler--Lagrange equation arising from the variation of $u$ at the central point is given by
\begin{gather}
\frac{\partial}{\partial u}\big(L\big(T_{i}^{-1}u,u,T_{i}^{-1}T_{j}u;\alpha_{i},\alpha_{j}\big)+L(u,T_{i}u,T_{j}u;\alpha_{i},\alpha_{j})\nonumber \\
\qquad {}+L\big(T_{j}^{-1}u,T_{i}T_{j}^{-1}u,u;\alpha_{i},\alpha_{j}\big)\big)= 0.\label{EL2d}
\end{gather}
The other elementary configurations in 3 dimensions are shown in Fig.~\ref{3d_configs}. All other configurations, including the ``flat configuration'' of Fig.~\ref{2d_config}, can be obtained as combinations of the latter three. (A similar statement to this effect has appeared in \cite{BolPetSur2013b}.) Note that in the last picture in Fig.~\ref{3d_configs} only two plaquettes contribute, because of the 3-point nature of the Lagrangians we are considering here.

Each of these surface configurations corresponds to a different Euler--Lagrange equation. Since all surfaces in the lattice can be obtained by combining these elementary configurations, the EL equation for any surface can be obtained by combining the Euler--Lagrange equations corresponding to the respective elementary configurations. Thus, we arrive at the following statement.

\begin{Theorem}
The following equations form the complete set of Euler--Lagrange equations describing the local critical points of the action functional \eqref{action}, with discrete Lagrangian $2$-form given by \eqref{Lagrangian}, with respect to infinitesimal variations $u(\boldsymbol{n})\to u(\boldsymbol{n})+\delta u(\boldsymbol{n})$ at each interior point of a quad-lattice surface, and with respect to local flips of the quadrilateral lattice
\begin{subequations}\label{eq:fundEL}
\begin{gather}
 \frac{\partial}{\partial u}\big(L(u,T_{i}u,T_{j}u;\alpha_{i},\alpha_{j})+L(u,T_{j}u,T_{k}u;\alpha_{j},\alpha_{k})+L(u,T_{k}u,T_{i}u;\alpha_{k},\alpha_{i})\big) = 0,
 \label{EL2}\\
 \frac{\partial}{\partial u}\big(L\big(T_{i}^{-1}u,u,T_{i}^{-1}T_{j}u;\alpha_{i},\alpha_{j}\big)-L(u,T_{j}u,T_{k}u;\alpha_{j},\alpha_{k}) \nonumber \\
 \qquad{}+L\big(T_{i}^{-1}u,T_{i}^{-1}T_{k}u,u;\alpha_{k},\alpha_{i}\big)\big) = 0,\label{EL3}\\
 \frac{\partial}{\partial u}\big(L\big(T_{j}^{-1}u,u,T_{j}^{-1}T_{k}u;\alpha_{j},\alpha_{k}\big)+L\big(T_{i}^{-1}u,T_{i}^{-1}T_{k}u,u;\alpha_{k},\alpha_{i}\big)\big) = 0, \label{EL4}
\end{gather}
for all $i,j,k\in I$, where $I$ is the index set labelling the lattice directions, together with the closure relation
\begin{gather} \label{EL5} \Delta_i L(u,T_ju,T_ku;\alpha_j,\alpha_k)+\Delta_j L(u,T_k u,T_i u;\alpha_k,\alpha_i)+ \Delta_k L(u,T_iu,T_ju;\alpha_i,\alpha_j)=0,
\end{gather}
\end{subequations}
where $\Delta_i=T_i-{\rm id}$ denotes the difference operator in the $i^{\rm th}$ direction.
\end{Theorem}

\begin{proof}Consider the action of a closed surface. The smallest non-trivial closed surface is a~cube, for which the action is
\begin{gather*}%\label{eq:cubeact}
 S[u;{\rm cube}] = L(u,T_{i}u,T_{j}u;\alpha_{i},\alpha_{j})+L(u,T_{j}u,T_{k}u;\alpha_{j},\alpha_{k})+L(u,T_{k}u,T_{i}u;\alpha_{k},\alpha_{i}) \\
\hphantom{S[u;{\rm cube}] =}{} -L(T_{k}u,T_{i}T_{k}u,T_{j}T_{k}u;\alpha_{i},\alpha_{j})-L(T_{i}u,T_{i}T_{j}u,T_{i}T_{k}u;\alpha_{j},\alpha_{k})\\
\hphantom{S[u;{\rm cube}] =}{} -L(T_{j}u,T_{j}T_{k}u,T_{i}T_{j}u;\alpha_{k},\alpha_{i}).
\end{gather*}

To derive elementary configurations we start with the action over the full oriented cube, in which triangles decorating the faces indicate the Lagrangian contributions of those faces to the action:

\begin{figure}[h]\centering
%%%%%%%%%%%%%%%%%%
%%% FLAT SPACE %%%
%%%%%%%%%%%%%%%%%%
\begin{tikzpicture}
	%%% Edit the following coordinate to change the shape of your
	%%% cuboid

	%% Vanishing points for perspective handling
	\coordinate (P1) at (-7cm,1.5cm); % left vanishing point (To pick)
	\coordinate (P2) at (8cm,1.5cm); % right vanishing point (To pick)

	%% (A1) and (A2) defines the 2 central points of the cuboid
	\coordinate (A1) at (0em,0cm); % central top point (To pick)
	\coordinate (A2) at (0em,-2cm); % central bottom point (To pick)

	%% (A3) to (A8) are computed given a unique parameter (or 2) .8
	% You can vary .8 from 0 to 1 to change perspective on left side
	\coordinate (A3) at ($(P1)!.8!(A2)$); % To pick for perspective
	\coordinate (A4) at ($(P1)!.8!(A1)$);

	% You can vary .8 from 0 to 1 to change perspective on right side
	\coordinate (A7) at ($(P2)!.7!(A2)$);
	\coordinate (A8) at ($(P2)!.7!(A1)$);

	%% Automatically compute the last 2 points with intersections
	\coordinate (A5) at
	 (intersection cs: first line={(A8) -- (P1)},
			 second line={(A4) -- (P2)});
	\coordinate (A6) at
	 (intersection cs: first line={(A7) -- (P1)},
			 second line={(A3) -- (P2)});

	%%% Depending of what you want to display, you can comment/edit
	%%% the following lines

	%% Possibly draw back faces

	\fill[green!50] (A2) -- (A3) -- (A6) -- cycle; % face 6
	%\node at (barycentric cs:A2=1,A3=1,A6=1,A7=1) {\tiny f6};
	
	\fill[gray!50] (A3) -- (A4) -- (A6) -- cycle; % face 3
	%\node at (barycentric cs:A3=1,A4=1,A5=1,A6=1) {\tiny f3};
	
	\fill[gray!30] (A5) -- (A6) -- (A7) -- cycle; % face 4
	%\node at (barycentric cs:A5=1,A6=1,A7=1,A8=1) {\tiny f4};
	
	\fill[green!30, opacity=0.4] (A3) -- (A2) -- (A4) -- cycle; % face 4
	%\node at (barycentric cs:A5=1,A6=1,A7=1,A8=1) {\tiny f4};

	\draw[thick,dashed] (A5) -- (A6);
	\draw[thick,dashed] (A2) -- (A6);
	\draw[thick] (A4) -- (A2);
 \draw[thick] (A5) -- (A1);
 \draw[thick,dashed] (A5) -- (A7);
 \draw[thick] (A1) -- (A7);
	\draw[thick,dashed] (A4) -- (A6);
	\draw[thick,dashed] (A3) -- (A6);
	\draw[thick,dashed] (A7) -- (A6);

	%% Possibly draw front faces

	\fill[green!30, fill opacity=0.5] (A1) -- (A7) -- (A2) -- cycle; % face 1
	% \node at (barycentric cs:A1=1,A8=1,A7=1,A2=1) {\tiny f1};
	\fill[gray!50,opacity=0.2] (A1) -- (A2) -- (A3) -- cycle; % f2
	%\node at (barycentric cs:A1=1,A2=1,A3=1,A4=1) {\tiny f2};
	\fill[gray!90,opacity=0.4] (A1) -- (A4) -- (A5) -- cycle; % f5
	%\node at (barycentric cs:A1=1,A4=1,A5=1,A8=1) {\tiny f5};

	%% Possibly draw front lines
	\draw[very thick] (A1) -- (A2);
	\draw[very thick] (A3) -- (A4);
	\draw[thick] (A7) -- (A8);
	\draw[thick] (A1) -- (A4);
	\draw[thick] (A1) -- (A8);
	\draw[very thick] (A2) -- (A3);
	\draw[very thick] (A2) -- (A7);
	\draw[very thick] (A4) -- (A5);
	\draw[thick] (A8) -- (A5);
	
	% Possibly draw points
	% (it can help you understand the cuboid structure)
	\foreach \i in {1,2,...,8}
	{
	 \draw[fill=black] (A\i) circle (0.20em);
	 node[above right] {\tiny \i};
	
	}
	 \node[below left] at (A1) {\small $T_iT_ju$} ;
	 \node[below right] at (A2) {\small $T_iu$} ;
	 \node[below left] at (A3) {\small $u$} ;
	 \node[above right] at (A7) {\small $T_iT_ku$} ;
	 \node[above right] at (A6) {\small $T_ku$} ;
	 \node[above left] at (A5) {\small $T_jT_ku$} ;
	 \node[above left] at (A4) {\small $T_ju$} ;
	 \node[above right] at (A8) {\small $T_iT_jT_ku$} ;

	% \draw[fill=black] (P1) circle (0.1em) node[below] {\tiny p1};
	% \draw[fill=black] (P2) circle (0.1em) node[below] {\tiny p2};
\end{tikzpicture}\vspace{-2mm}
\caption{Decorated cube.}\label{fullcubeflatspace}
\end{figure}
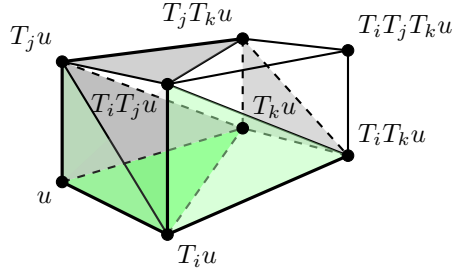

We now require variations of the action with respect to the dependent variables to be zero at each vertex of the closed decorated surface. This gives rise to local configurations of the three types given in Fig.~\ref{3d_configs}. These lead then to the following three types of EL equations respectively
\begin{subequations}\label{abc}
\begin{gather}
 \frac{\partial}{\partial u}\big(L(u,T_{i}u,T_{j}u;\alpha_{i},\alpha_{j})+L(u,T_{j}u,T_{k}u;\alpha_{j},\alpha_{k})+L(u,T_{k}u,T_{i}u;\alpha_{k},\alpha_{i})\big) = 0, \label{a}\\
 \frac{\partial}{\partial T_{i}u}\big(L(u,T_{i}u,T_{j}u;\alpha_{i},\alpha_{j})-L(T_{i}u,T_{i}T_{j}u,T_{i}T_{k}u;\alpha_{j},\alpha_{k})\nonumber\\
 \qquad{} +L(u,T_{k}u,T_{i}u;\alpha_{k},\alpha_{i})\big) = 0, \label{b}\\
 \frac{\partial}{\partial T_{j}T_{k}u}\big(L(T_{k}u,T_{i}T_{k}u,T_{j}T_{k}u;\alpha_{i},\alpha_{j})+L(T_{j}u,T_{j}T_{k}u,T_{i}T_{j}u;\alpha_{k},\alpha_{i})\big) = 0, \label{c}
\end{gather}
\end{subequations}
and cyclic permutations. Shifting these in the lattice we see that they are equivalent to \eqref{EL2}--\eqref{EL4}. Any closed quad-surface can be constructed from cubes, so at least away from any boundary all possible Euler--Lagrange equations are consequences of \eqref{abc}. Furthermore, the closure relation~\eqref{EL5} guarantees that the system of EL equations~\eqref{eq:fundEL} continues to hold true under elementary lattice flips, and hence under local deformations of the quad-lattice.
\end{proof}

We will regard the system \eqref{eq:fundEL}, given by a set of linear partial difference equations for the function~$L$, to be the fundamental system of (extended) EL equations, that defines not only the dynamics but determines also the Lagrangian 2-form $\mathcal{L}_{ij}$ itself.

\subsection{Consequences of the system of Euler--Lagrange equations}\label{section2.3}

As in the previous subsection, we consider actions which are the sum of 3-point Lagrangians $L(u,T_{i}u,T_{j}u;\alpha_{i},\alpha_{j})$, where the Lagrangians are antisymmetric with respect to the interchange of the lattice directions, so that the equations~\eqref{eq:fundEL} hold. The one further assumption we will make is that we may choose initial conditions $u$, $T_{i}u$, $T_{j}u$, $T_{k}u$ independently and arbitrarily. We empasize that these generalised Euler--Lagrange equations form a set of linear PDEs for the Lagrange function $L(u,v,w;\alpha,\beta)$, which (apart from the arguments written), may still depend on the lattice variables $\boldsymbol{n}$, leading possible to non-autonomous equations. We will treat this system as a set of functional equations with regard to the arguments where the arbitrariness of the choice of lattice directions (i.e., the dependence on the labels $i$, $j$, $k$ referring associated with the lattice shifts), and hence the property of MDC, plays a central role.

If we impose that the action remains invariant under deformations of the surface, then it is independent of the surface~\cite{LobNij2009}, and all of these equations must hold simultaneously. Note that~\eqref{EL2d} is a consequence of \eqref{EL2}--\eqref{EL4} and their cyclic permutations.

\begin{Theorem}Suppose $u$, $T_{i}u$, $T_{j}u$, $T_{k}u$ are independent and can be chosen arbitrarily in all directions $i,j,k\in I$. The Euler--Lagrange equation \eqref{EL2} implies that the skew symmetric Lagrangian $L(u,T_{i}u,T_{j}u;\alpha_{i},\alpha_{j})$ has the form
\begin{gather}\label{3ptform}
 L(u,T_{i}u,T_{j}u;\alpha_{i},\alpha_{j}) = A(u,T_{i}u;\alpha_{i})-A(u,T_{j}u;\alpha_{j})+C(T_{i}u,T_{j}u;\alpha_{i},\alpha_{j}),
\end{gather}
for some functions $A$ and $C$, with $C$ skew symmetric,
$C(T_{i}u,T_{j}u;\alpha_{i},\alpha_{j})=C(T_{j}u,T_{i}u;\alpha_{j},\alpha_{i})$.
\end{Theorem}
\begin{proof}Consider equation \eqref{EL2} in a single rhomboid. If $u$, $T_{i}u$, $T_{j}u$, $T_{k}u$ for all directions $i,j,k\in I$ can be chosen arbitrarily, then writing
\begin{gather*}
 l(u,T_{i}u,T_{j}u;\alpha_{i},\alpha_{j}) = \frac{\partial}{\partial u}L(u,T_{i}u,T_{j}u;\alpha_{i},\alpha_{j}),
\end{gather*}
we must have
\begin{gather}
 \frac{\partial}{\partial T_{i}u}\big(l(u,T_{i}u,T_{j}u;\alpha_{i},\alpha_{j})+l(u,T_{k}u,T_{i}u;\alpha_{k},\alpha_{i})\big) = 0,\nonumber\\
 \qquad\Rightarrow \quad l(u,T_{i}u,T_{j}u;\alpha_{i},\alpha_{j}) = a(u,T_{i}u;\alpha_{i})+b(u,T_{j}u;\alpha_{i},\alpha_{j}),\label{eq:9}
\end{gather}
for some functions $a$, $b$. Equation~\eqref{eq:9}, together with the various permutations of the lattice directions, yields
\begin{gather*}
 l(u,T_{i}u,T_{j}u;\alpha_{i},\alpha_{j}) = a(u,T_{i}u;\alpha_{i})-a(u,T_{j}u;\alpha_{j})+d(u;\alpha_i,\alpha_j),
\end{gather*}
for some function $d$, which is skew-symmetric in $\alpha_i$, $\alpha_j$. Thus, setting $\partial A(u,T_{i}u;\alpha_{i})/\partial u=a(u,T_{i}u;\alpha_{i})$, $\partial D(u;\alpha_{i},\alpha_{j})/\partial u=d(u;\alpha_{i},\alpha_{j})$, we have
\begin{gather*}
 L(u,T_{i}u,T_{j}u;\alpha_{i},\alpha_{j}) = A(u,T_{i}u;\alpha_{i})-A(u,T_{j}u;\alpha_{j}) + C(T_{i}u,T_{j}u;\alpha_{i},\alpha_{j})+D(u;\alpha_i,\alpha_j),
\end{gather*}
for some skew symmetric functions $C$, and $D$. From the closure relation \eqref{EL5}, together with~\eqref{EL2}, however, it follows that $D$ must be independent of $u$, and thus it can be absorbed in the function $C$, leading to~\eqref{3ptform}. Note that since $L(u,T_{i}u,T_{j}u;\alpha_{i},\alpha_{j})$ is antisymmetric under the interchange of lattice directions~$i$, $j$, then the same must be true of $C(T_{i}u,T_{j}u;\alpha_{i},\alpha_{j})$.
\end{proof}

Note that, since the direction analysis in the proof involves only a single rhomboid in the lattice, the functions $A$ and $C$ in \eqref{3ptform} may, apart from the written arguments, still depend on the lattice site $\boldsymbol{n}$, but in a way that is independent of the lattice direction in the sense that they obey $(T_iA)(u,v,\dots)=(T_jA)(u,v,\dots)$ for all $i,j\in I$, where the shift operators $T_i$, $T_j$ are not acting on
the arguments of the function.

From the perspective presented here the EL equations \eqref{eq:fundEL} are fundamental, while the corresponding quad equations are in a sense a mere corollary. The connection between the former and quad equations, which arise as `integrated' objects, in fact as conservation laws, is given by the following\footnote{A similar result in the literature, but in a~different perspective, can be found in \cite[Theorem~9.3]{BolPetSur2015a}.}:

\begin{Theorem} For skew symmetric Lagrangians of the general $3$-point form \eqref{3ptform} the following quadrilateral relation
\begin{gather}\label{quadrel}
 \frac{\partial}{\partial u}\big(L\big(T_{i}^{-1}u,u,T_{i}^{-1}T_{j}u;\alpha_{i},\alpha_{j}\big)\big) - \frac{\partial}{\partial u}\big(A(u,T_{j}u;\alpha_{j})\big)=h(u),
\end{gather}
constitutes a conservation law for the Euler--Lagrange equations \eqref{eq:fundEL}, under the assumption that the equations hold in any triplet of lattice directions, indicated by the labels $i,j,k\in I$, of a~lattice of arbitrary number $(\geq 3)$ of lattice directions, and where the function $h(u)=h(u;\boldsymbol{n})$ is independent of the lattice direction $($i.e., it obeys $h(u;T_i\boldsymbol{n})=h(u;T_j\boldsymbol{n})$ for all $i,j\in I)$.
\end{Theorem}

\begin{proof}Substituting \eqref{3ptform} into equations \eqref{EL3} and \eqref{EL4} gives
\begin{gather*}
 \frac{\partial}{\partial u}\big({-}A\big(T_{i}^{-1}u,T_{i}^{-1}T_{j}u;\alpha_{j}\big)-A(u,T_{j}u;\alpha_{j})+A(u,T_{k}u;\alpha_{k})+A\big(T_{i}^{-1}u,T_{i}^{-1}T_{k}u;\alpha_{k}\big) \\
 \qquad {}+C\big(u,T_{i}^{-1}T_{j}u;\alpha_{i},\alpha_{j}\big)-C\big(u,T_{i}^{-1}T_{k}u;\alpha_{i},\alpha_{k}\big)\big) = 0,\label{EL3b}\\
 \frac{\partial}{\partial u}\big(A\big(T_{j}^{-1}u,u;\alpha_{j}\big)-A\big(T_{j}^{-1}u,T_{j}^{-1}T_{k}u;\alpha_{k}\big)+A\big(T_{i}^{-1}u,T_{i}^{-1}T_{k}u;\alpha_{k}\big)-A\big(T_{i}^{-1}u,u;\alpha_{i}\big) \\
\qquad {}+C\big(u,T_{j}^{-1}T_{k}u;\alpha_{j},\alpha_{k}\big)-C\big(u,T_{i}^{-1}T_{k}u;\alpha_{i},\alpha_{k}\big)\big) = 0,\label{EL4b}
\end{gather*}
where we have already cancelled some of the terms (provided we assume that $T_{j}u$ and $T_{k}u$ can be independently chosen, so that they don't depend on the choice of~$u$). We can rewrite these in a suggestive way, isolating the dependence on particular lattice directions
\begin{gather*}
 \frac{\partial}{\partial u}\big(A(u,T_{j}u;\alpha_{j})+A\big(T_{i}^{-1}u,T_{i}^{-1}T_{j}u;\alpha_{j}\big)-C\big(u,T_{i}^{-1}T_{j}u;\alpha_{i},\alpha_{j}\big)\big) \\
 \qquad {}= \frac{\partial}{\partial u}\big(A(u,T_{k}u;\alpha_{k})+A\big(T_{i}^{-1}u,T_{i}^{-1}T_{k}u;\alpha_{k}\big)-C\big(u,T_{i}^{-1}T_{k}u;\alpha_{i},\alpha_{k}\big)\big),\\ %\label{EL3c}\\
 \frac{\partial}{\partial u}\big(A\big(T_{j}^{-1}u,u;\alpha_{j}\big)-A\big(T_{j}^{-1}u,T_{j}^{-1}T_{k}u;\alpha_{k}\big)+C\big(u,T_{j}^{-1}T_{k}u;\alpha_{j},\alpha_{k}\big)\big) \\
 \qquad{} = \frac{\partial}{\partial u}\big(A\big(T_{i}^{-1}u,u;\alpha_{i}\big)-A\big(T_{i}^{-1}u,T_{i}^{-1}T_{k}u;\alpha_{k}\big)+C\big(u,T_{i}^{-1}T_{k}u;\alpha_{i},\alpha_{k}\big)\big),
\end{gather*}
and of course this must be true for all $i$, $j$, $k$. Thus we must have
\begin{subequations}\label{eq:FG}
\begin{gather}
 \frac{\partial}{\partial u}\big(A(u,T_{j}u;\alpha_{j})+A\big(T_{i}^{-1}u,T_{i}^{-1}T_{j}u;\alpha_{j}\big)-C\big(u,T_{i}^{-1}T_{j}u;\alpha_{i},\alpha_{j}\big)\big) = f_i ,\\
 \frac{\partial}{\partial u}\big(A\big(T_{i}^{-1}u,u;\alpha_{i}\big)-A\big(T_{i}^{-1}u,T_{i}^{-1}T_{j}u;\alpha_{j}\big)+C\big(u,T_{i}^{-1}T_{j}u;\alpha_{i},\alpha_{j}\big)\big) = g_j,
\end{gather}
\end{subequations}
for some quantities $f_i$, $g_j$ in which only one of the lattice directions, indicated by their index~$i$,~$j$ respectively, is distinguished\footnote{A particular choice of such quantities is where they are functions of $u$ and the forward and backward shifts in the corresponding lattice direction, i.e., $f_i=f\big(u,T_iu,T_i^{-1}u,\dots\big)$, and $g_j=g\big(u,T_ju,T_j^{-1}u,\dots\big)$, but the argument in the proof is not depending on such a choice.}. Adding the relations \eqref{eq:FG} together, we find that
\begin{gather*} f_i+g_j=\frac{\partial}{\partial u}\big(A\big(T_{i}^{-1}u,u;\alpha_{i}\big)+A(u,T_{j}u;\alpha_{j})\big), \end{gather*}
for any two direction labels $i,j\in I$, and, hence, we deduce that
\begin{subequations}\label{eq:fg}
\begin{gather}
 f_i = \frac{\partial}{\partial u}\big(A\big(T_{i}^{-1}u,u;\alpha_{i}\big)\big)-h, \label{eq:fga}\\
 g_j = \frac{\partial}{\partial u}\big(A(u,T_{j}u;\alpha_{j})\big)+h, \label{eq:fgb}
\end{gather}
\end{subequations}
for some quantity $h$ in which none of the lattice directions is distinguished. In what follows we will chose this quantity to be a function of $u$ alone, i.e., $h=h(u)$, within the chosen rhomboid, while its dependence on $\boldsymbol{n}$ is such that $h(u;T_i\boldsymbol{n})=h(u;T_j\boldsymbol{n})$. Reinserting~\eqref{eq:fg} into \eqref{eq:FG}, we obtain from both~\eqref{eq:fga} and~\eqref{eq:fgb} one and the same relation on a single quad, namely
\begin{gather*}
 \frac{\partial}{\partial u}\big(A\big(T_{i}^{-1}u,u;\alpha_{i}\big)-A\big(T_{i}^{-1}u,T_{i}^{-1}T_{j}u;\alpha_{j}\big)+C\big(u,T_{i}^{-1}T_{j}u;\alpha_{i},\alpha_{j}\big)\big)\\
 \qquad{} = \frac{\partial}{\partial u}\big(A(u,T_{j}u;\alpha_{j})\big)+h,
\end{gather*}
which is the stated conservation law \eqref{quadrel}. It is straightforward to check that~\eqref{3ptform} with~\eqref{quadrel} are enough to satisfy all Euler--Lagrange equations \eqref{EL2}--\eqref{EL4}.
\end{proof}

We point out that in the search for solutions of the EL equations \eqref{EL2}--\eqref{EL4}, of the form~\eqref{3ptform} where $A$ is yet unspecified, and assuming that~$h$ is a function of~$u$ at a given vertex only (and functionally independent of its shifts $T_iu$, $T_ju$, $T_ku$), the function $h(u)$ can be absorbed into the $A$, and hence does not need to be considered separately. Furthermore, the analysis only takes into account the directional dependence in the EL equations (i.e., the dependence on the direction labels~$i$, $j$, $k$), and does not specify the dependence on the lattice variables~$\boldsymbol{n}$. Thus, the quantity~$h$, and hence the function~$A$, can still depend on~$\boldsymbol{n}$, but only in a way which doesn't distinguish any specific lattice direction. Thus, the form of the quad-equation \eqref{quadrel} takes either of the following equivalent form
\begin{subequations}
\begin{gather}
 \frac{\partial}{\partial T_{i}u}\big\{A(u,T_{i}u;\alpha_{i})-A(T_{i}u,T_{i}T_{j}u;\alpha_{j})+C(T_{i}u,T_{j}u;\alpha_{i},\alpha_{j})\big\} = 0,\label{ieqn}\\
 \frac{\partial}{\partial T_{j}u}\big\{A(T_{j}u,T_{i}T_{j}u;\alpha_{i})-A(u,T_{j}u;\alpha_{j})+C(T_{i}u,T_{j}u;\alpha_{i},\alpha_{j})\big\} = 0,\label{jeqn}
\end{gather}
\end{subequations}
where $A$ remains to be identified. If the resulting quad equation obeys the MDC property, then the resulting Lagrangian obeys by construction the closure property \eqref{EL5}.

In what follows we will illustrate the analysis by considering specific examples of 3-point Lagrangians of the form \eqref{3ptform} leading via \eqref{quadrel} to some well-known quad-equations that possess the MDC property.
In doing this we will start with a given form of the 3-point Lagrangian, and identify the form of the function~$A$.

\subsection{Example: H1}
In the example of the quad-equation called H1 in \cite{AdlBobSur2003}, which is the lattice potential KdV equation, the Lagrangian (which was first given in~\cite{CapNijPap1991}) evaluated on a plaquette in the $(i,j)$-direction has the form
\begin{gather}\label{H1Lagr}
 L(u,T_{i}u,T_{j}u;\alpha_{i},\alpha_{j}) = (T_{i}u-T_{j}u)u-(\alpha_{i}-\alpha_{j})\ln(T_{i}u-T_{j}u).
\end{gather}
The usual Euler--Lagrange equation \eqref{EL2d} coming from an action on a flat 2-d surface is
\begin{subequations}
\begin{gather}
\frac{\partial}{\partial u}\big(\big(T_{i}u-T_{j}u+T_{i}^{-1}u-T_{j}^{-1}u\big)u-(\alpha_{i}-\alpha_{j})\ln\big(u-T_{i}^{-1}T_{j}u\big) \nonumber \\
\qquad\qquad {}-(\alpha_{i}-\alpha_{j})\ln\big(T_{i}T_{j}^{-1}u-u\big)\big)=0, \nonumber \\
\qquad \Rightarrow \quad T_{i}u-T_{j}u+T_{i}^{-1}u-T_{j}^{-1}u-\frac{\alpha_{i}-\alpha_{j}}{u-T_{i}^{-1}T_{j}u}+\frac{\alpha_{i}-\alpha_{j}}{T_{i}T_{j}^{-1}u-u}=0,
\label{H1a}
\end{gather}
which consists of 2 shifted copies of H1 lying on a 7-point configuration, i.e., a difference consequence of H1. The Euler--Lagrange equations on non-flat surfaces \eqref{EL2}--\eqref{EL4} yield the two relations
\begin{gather}
\frac{\partial}{\partial u}\big(u(-T_{j}u+T_{k}u)-(\alpha_{i}-\alpha_{j})\ln\big(u-T_{i}^{-1}T_{j}u\big)-(\alpha_{k}-\alpha_{i})\ln\big(T_{i}^{-1}T_{k}u-u\big)\big)= 0,\nonumber\\
\qquad \Rightarrow \quad -T_{j}u+T_{k}u-\frac{\alpha_{i}-\alpha_{j}}{u-T_{i}^{-1}T_{j}u}+\frac{\alpha_{k}-\alpha_{i}}{T_{i}^{-1}T_{k}u-u}= 0,\label{H1c} \\
\frac{\partial}{\partial u}\big(u\big(T_{j}^{-1}u-T_{i}^{-1}u\big)-(\alpha_{j}-\alpha_{k})\ln\big(u-T_{j}^{-1}T_{k}u\big)-(\alpha_{k}-\alpha_{i})\ln\big(T_{i}^{-1}T_{k}u-u\big)\big)=
 0 ,\nonumber\\
\qquad \Rightarrow \quad T_{j}^{-1}u-T_{i}^{-1}u-\frac{\alpha_{j}-\alpha_{k}}{u-T_{j}^{-1}T_{k}u}+\frac{\alpha_{k}-\alpha_{i}}{T_{i}^{-1}T_{k}u-u} = 0,\label{H1d}
\end{gather}
\end{subequations}
from \eqref{EL3} and \eqref{EL4} respectively, while \eqref{EL2} in this case is trivial. In fact, \eqref{H1d} is a~consequence of \eqref{H1c} and its copies under permutation of lattice directions.

Turning now to equation \eqref{quadrel}, we need to identify the function $A$, where we can set $h=0$ as $A$ is yet unspecified. This yields
\begin{gather}
 T_{i}^{-1}u-\frac{\alpha_{i}-\alpha_{j}}{u-T_{i}^{-1}T_{j}u} = \frac{\partial}{\partial u}\big(A(u,T_{j}u;\alpha_{j})\big)\nonumber\\
 \qquad \Rightarrow\quad u-\frac{\alpha_{i}-\alpha_{j}}{T_{i}u-T_{j}u} = \frac{\partial}{\partial T_{i}u}\big(A(T_{i}u,T_{i}T_{j}u;\alpha_{j})\big). \label{quadrelH1}
\end{gather}
Of course, by swapping the lattice directions, we also get
\begin{gather}
 u-\frac{\alpha_{i}-\alpha_{j}}{T_{i}u-T_{j}u} = \frac{\partial}{\partial T_{j}u}\big(A(T_{j}u,T_{i}T_{j}u;\alpha_{i})\big).\label{quadrelH1'}
\end{gather}
Combining \eqref{quadrelH1} and \eqref{quadrelH1'} we get
\begin{gather*}
 \frac{\partial}{\partial T_{i}u}\big(A(T_{i}u,T_{i}T_{j}u;\alpha_{j})\big) = \frac{\partial}{\partial T_{j}u}\big(A(T_{j}u,T_{i}T_{j}u;\alpha_{i})\big)= f(T_{i}T_{j}u),
\end{gather*}
for some function $f(u)$ which is independent of the lattice directions in the sense as explained in Section~\ref{section2.3} (which does not preclude $f$ to depend on the lattice variables $\boldsymbol{n}$, but only in a way that does not distinguish any of the lattice direction labels). This implies the following form of the function~$A$
\begin{gather*}
 A(u,T_{j}u;\alpha_{j}) = uf(T_{j}u)+g(T_{j}u;\alpha_{j}).
\end{gather*}

In order to identify the resulting Lagrangian \eqref{3ptform} with \eqref{H1Lagr} we set
\begin{gather*}
 A(u,T_{i}u;\alpha_{i})-A(u,T_{j}u;\alpha_{j}) = (T_{i}u-T_{j}u)u,
\end{gather*}
and so we must have $f(x)=x+\lambda$, $g(x,\alpha)=\mu$, with $\lambda=\lambda(\boldsymbol{n})$ and $\mu=\mu(\boldsymbol{n})$ constant with regard to its dependence on the field variable $u$, and in which none of the lattice directions is distinguished (implying that any dependence on the lattice site $\boldsymbol{n}$ is such that $T_i\lambda=T_j\lambda$, and similar for $\mu$), leading to
\begin{gather*}
 A(u,T_{i}u;\alpha_{i}) = (T_{i}u + \lambda)u+\mu.
\end{gather*}
Then the quad relation \eqref{quadrel} adopts the form
\begin{gather}\label{eq:H1}
 \big(u-\bar{\lambda}-T_{i}T_{j}u\big)(T_{i}u-T_{j}u)-\alpha_{i}+\alpha_{j}=0,
\end{gather}
which is consistent around a cube for arbitrary direction independent: $\bar{\lambda}:=T_i\lambda=T_j\lambda$. Note that the latter quantity cancels out in the usual Euler--Lagrange equation \eqref{H1a}, and that it can be absorbed into the function $u$ by a point transformation. Hence, we can set it equal to zero w.l.o.g., in which case equation~\eqref{eq:H1} coincides with the H1 equation (also known as lattice potential KdV equation) in the list of~\cite{AdlBobSur2003}. Thus, we retrieve from the full multiform EL system~\eqref{eq:fundEL} the quad equation, but the latter equation depends on the solutions of the EL equation.

\subsection{Example: H3}
The Lagrangian evaluated on a plaquette in the $(i,j)$-direction has the form
\begin{gather*}
 L(u,T_{i}u,T_{j}u;\alpha_{i},\alpha_{j}) = \ln\big(\alpha_{i}^2\big)\ln{u}-{\rm Li}_{2}\left(-\frac{uT_{i}u}{\alpha_{i}\delta}\right)-\ln\big(\alpha_{j}^2\big)\ln{u}+{\rm Li}_{2}\left(-\frac{uT_{j}u}{\alpha_{j}\delta}\right)\nonumber\\
\hphantom{L(u,T_{i}u,T_{j}u;\alpha_{i},\alpha_{j}) =}{} +{\rm Li}_{2}\left(\frac{\alpha_{j}T_{i}u}{\alpha_{i}T_{j}u}\right)-{\rm Li}_{2}\left(\frac{\alpha_{i}T_{i}u}{\alpha_{j}T_{j}u}\right)+\ln\big(\alpha_{j}^2\big)\ln\left(\frac{T_{i}u}{T_{j}u}\right),
\end{gather*}
where $\delta$ is an arbitrary constant parameter. Here ${\rm Li}_{2}(x)$ is the well known dilogarithm function, defined by
\begin{gather*} {\rm Li}_2(x) = -\int^x_0\,x^{-1}\ln(1-x)\,{\rm d}x. \end{gather*}
Up to an as yet unspecified function $f(u)$ we can identify
\begin{subequations} \label{eq:H3AC}
\begin{gather}
 A(u,T_{i}u;\alpha_{i}) = \ln\big(\alpha_{i}^2\big)\ln{u}-{\rm Li}_{2}\left(-\frac{uT_{i}u}{\alpha_{i}\delta}\right) +f(u) , \label{eq:H3A} \\
 C(T_{i}u,T_{j}u;\alpha_{i},\alpha_{j}) = {\rm Li}_{2}\left(\frac{\alpha_{j}T_{i}u}{\alpha_{i}T_{j}u}\right)-{\rm Li}_{2}\left(\frac{\alpha_{i}T_{i}u}{\alpha_{j}T_{j}u}\right)+\ln\big(\alpha_{j}^2\big)\ln\left(\frac{T_{i}u}{T_{j}u}\right). \label{eq:H3C}
\end{gather}
\end{subequations}

Note that the quantity $C$ of \eqref{eq:H3C} is antisymmetric as a consequence of the property of the dilogarithm
\begin{gather*} {\rm Li}_2(x)+{\rm Li}_2(1/x)=-\frac{\pi^2}{6}-\frac{1}{2} (\ln(-x) )^2. \end{gather*}

The equation \eqref{ieqn} is then
\begin{gather*}
 \frac{\partial}{\partial T_{i}u}\left\{\ln\big(\alpha_{i}^2\big)\ln{u}-{\rm Li}_{2}\left(-\frac{uT_{i}u}{\alpha_{i}\delta}\right) +f(u)
 -\ln\big(\alpha_{j}^2\big)\ln(T_{i}u)+{\rm Li}_{2}\left(-\frac{(T_{i}u)T_{i}T_{j}u}{\alpha_{j}\delta}\right) \right.\\
 \left.\qquad {}-f(T_{i}u) +{\rm Li}_{2}\left(\frac{\alpha_{j}T_{i}u}{\alpha_{i}T_{j}u}\right)-{\rm Li}_{2}\left(\frac{\alpha_{i}T_{i}u}{\alpha_{j}T_{j}u}\right)+\ln(\alpha_{j}^2)\ln\left(\frac{T_{i}u}{T_{j}u}\right)\right\} = 0 \\
 \Rightarrow \quad \frac{1}{T_{i}u}\ln\left(1+\frac{u(T_{i}u)}{\alpha_{i}\delta}\right)-\frac{1}{T_{i}u}\ln\big(\alpha_{j}^2\big)-\frac{1}{T_{i}u}\ln\left(1+\frac{(T_{i}u)T_{i}T_{j}u}{\alpha_{j}\delta}\right)-f'(T_{i}u) \\
 \qquad {} -\frac{1}{T_{i}u}\ln\left(1-\frac{\alpha_{j}T_{i}u}{\alpha_{i}T_{j}u}\right)+\frac{1}{T_{i}u}\ln\left(1-\frac{\alpha_{i}T_{i}u}{\alpha_{j}T_{j}u}\right)+\frac{1}{T_{i}u}\ln\big(\alpha_{j}^2\big) = 0 \\
 \Rightarrow \quad \frac{1}{T_{i}u}\ln\left(\frac{u(T_{i}u)+\alpha_{i}\delta}{(T_{i}u)T_{i}T_{j}u+\alpha_{j}\delta}\cdot\frac{\alpha_{j}T_{j}u-\alpha_{i}T_{i}u}{\alpha_{i}T_{j}u-\alpha_{j}T_{i}u}\right)-f'(T_{i}u) = 0,
\end{gather*}
where $f'(z)=df/dz$.
If we define
\begin{gather*}
 t_{i} = \exp\big\{(T_{i}u)f'(T_{i}u)\big\},
\end{gather*}
then the exponentiation of the latter result yields
\begin{gather} \label{H3leg}
\frac{\alpha_i\delta+u(T_iu)}{\alpha_j\delta+(T_iu)T_iT_ju}=t_i \frac{\alpha_iT_ju-\alpha_jT_iu}{\alpha_jT_ju-\alpha_iT_iu},
\end{gather}
which is a slight generalization (due to the function $t_i$) of a particular 3-leg form of the H3 equation of \cite{AdlBobSur2003} (i.e., up to a symmetry of the square). To determine the function $t_i$, note that the equation \eqref{H3leg} can also be cast in the form
\begin{gather*}
 \alpha_{i}(uT_{i}u+t_{i}(T_{j}u)T_{i}T_{j}u)-\alpha_{j}(u(T_{j}u)+t_{i}(T_{i}u)T_{i}T_{j}u)+\delta\big(\alpha_{i}^2-t_{i}\alpha_{j}^2\big)\\
 \qquad{} +\alpha_{i}\alpha_{j}\delta\frac{T_{j}u}{T_{i}u}(t_{i}-1) = 0.
\end{gather*}
Note from \eqref{jeqn} that we also have the equation
\begin{gather*}
 \alpha_{i}(u(T_{i}u)+t_{j}(T_{j}u)T_{i}T_{j}u)-\alpha_{j}(u(T_{j}u)+t_{j}(T_{i}u)T_{i}T_{j}u)-\delta\big(\alpha_{j}^2-t_{j}\alpha_{i}^2\big)\\
 \qquad{} -\alpha_{i}\alpha_{j}\delta\frac{T_{i}u}{T_{j}u}(t_{j}-1) = 0,
\end{gather*}
and so we must have
\begin{gather*}
 (\alpha_{i}T_{j}u-\alpha_{j}T_{i}u)(T_{i}T_{j}u)(t_{i}-t_{j})\\
 \qquad{} +\delta\left[\alpha_{j}(t_{i}-1)\left(\alpha_{i}\frac{T_{j}u}{T_{i}u}-\alpha_{j}\right)+\alpha_{i}(t_{j}-1)\left(\alpha_{j}\frac{T_{i}u}{T_{j}u}-\alpha_{i}\right)\right]=0.
\end{gather*}
Therefore $t_{i}=t_{j}$, and if $\delta\neq 0$ then $t_{i}=t_{j}=1$, and we recover the H3 equation of~\cite{AdlBobSur2003}, in the form
\begin{gather*}
 \alpha_{i}\bigl(u(T_{i}u)+(T_{j}u)T_{i}T_{j}u\bigr)-\alpha_{j}\bigl(u(T_{j}u)+(T_{i}u)T_{i}T_{j}u\bigr)= \delta\big(\alpha_{j}^2-\alpha_{i}^2\big).
\end{gather*}
If, on the other hand, $\delta=0$ we have a little more freedom, and we can let $t_{i}=t_{j}=t$ for some arbitrary direction independent quantity $t$ which may depend on the lattice sites $\boldsymbol{n}$ but subject to $T_it=T_jt$ for all direction labels $i,j\in I$. In that case, the equation reads
\begin{gather*}
 \alpha_{i}\big(u(T_{i}u)+t(T_{j}u)T_{i}T_{j}u\big)-\alpha_{j}\big(u(T_{j}u)+t(T_{i}u)T_{i}T_{j}u\big)=0,
\end{gather*}
and this equation, which is equivalent to $({\rm H3})_{\delta=0}$, up to a point transformation, is also consistent around the cube. We mention that the possible explicit dependence on the lattice variables $\boldsymbol{n}$ (i.e., non-autonomicity) appearing in the form of the quad equations is not a new feature, and was studied in several places in the literature, cf., e.g.,~\cite{SRH2007}.

\subsection{Quadratic Lagrangians}

In the case of quadratic Lagrangians a full solution of the extended system of EL equations~\eqref{eq:fundEL}, can be found as follows.

Let us consider the general homogeneous quadratic Lagrangian 2-form component, which must be of the form~\eqref{3ptform}, namely
\begin{gather*} L_{i,j}(u,T_iu,T_ju)=A_i(u,T_iu)-A_j(u,T_ju)+B_{i,j}(T_iu,T_ju), \end{gather*}
by setting
\begin{gather*} A_i(u,T_iu) = \tfrac{1}{2}a_iu^2+a_i'uT_iu+\tfrac{1}{2}a''_i(T_iu)^2, \\
 B_{i,j}(T_iu,T_ju) =\tfrac{1}{2}b_{ij}(T_iu)^2-\tfrac{1}{2}b_{ji}(T_ju)^2+b'_{ij}(T_iu)T_ju,
\end{gather*}
where $b'_{ji}=-b'_{ij}$. Applying equation~\eqref{quadrel} in the form
\begin{gather*}
 \frac{\partial}{\partial T_{i}u}\big( L_{ij}(u,T_iu,T_ju)\big) = \frac{\partial}{\partial T_{i}u}\big( A_j(T_iu,T_iT_ju)\big),\label{Jeqn}
\end{gather*}
(which holds for all directions $i$, $j$) we obtain the linear quad-equation
\begin{gather*}
 a'_i u+(a''_i+b_{ij}-a_j)T_iu+b'_{ij}T_ju=a'_jT_iT_ju .
\end{gather*}
Since these hold for arbitrary $i,j$-labels we obtain the conditions
\begin{gather*} {a'}_i^2={a'}_j^2 , \qquad {\rm and}\qquad (a_i''+b_{ij}-a_j)a_i'=a_j'b'_{ji} . \end{gather*}
Setting (in accordance with the requirement of covariance) $a'_i=a'_j=:a'$ and implementing the other condition we get the quad-equation $T_iT_ju=u+\tfrac{1}{a'}b'_{ij}(T_ju-T_iu)$, with Lagrangian
\begin{gather*} L_{i,j}= \tfrac{1}{2}(a_i-a_j)u^2+a'u(T_iu-T_ju)+\tfrac{1}{2}a_j(T_iu)^2-\tfrac{1}{2}a_i(T_ju)^2 -\tfrac{1}{2}b'_{ij}(T_iu-T_ju)^2, \end{gather*}
where the terms with $a_i$ can be removed w.l.o.g.\ (since they form an exact discrete 2-form), and we can fix the overall constant by setting $a'=1$. The closure relation \eqref{EL5} leads to the functional relation
\begin{gather*} b'_{ij}(b'_{ik}-b'_{jk})=1-b'_{ik}b'_{jk}\quad \Rightarrow \quad b'_{ij}=(1-P_iP_j)/(P_i-P_j), \end{gather*}
in terms of a new lattice parameter $P_i:=b'_{i,k_0}$ for some fixed $k_0$. Thus, we obtain the unique (up to a multiplicative constant) Lagrangian 2-form
\begin{gather*}
L(u,T_iu,T_ju;\alpha_i,\alpha_j)=u(T_iu-T_ju)-\frac{1}{2} \frac{1-P_iP_j}{P_i-P_j}(T_iu-T_ju)^2.
\end{gather*}
More specifically, setting $b'_{ik_0}=:1+\epsilon \alpha_i$, and taking the limit $\epsilon\to 0$, the quadratic Lagrangian adopts the form
\begin{gather*}
L(u,T_iu,T_ju;\alpha_i,\alpha_j)=u(T_iu-T_ju)+\frac{1}{2} \frac{\alpha_i+\alpha_j}{\alpha_i-\alpha_j}(T_iu-T_ju)^2,
\end{gather*}
which produces the family of multidimensional consistent quad-equations
\begin{gather*}
(\alpha_i+\alpha_j)(T_iu-T_ju)=(\alpha_i-\alpha_j)(T_iT_ju-u).
\end{gather*}
Similar Lagrangian structures were considered in \cite{BobSur15}.

\section{Discrete 3-dimensional systems}\label{section_3d}

So far the only instance of a Lagrangian structure for a fully discrete 3-dimensional integrable equation was, to our knowledge, given in~\cite{LobNijQui2009}, namely for the Hirota bilinear KP equation. In that paper a description of the system was given in terms of a discrete Lagrangian 3-form structure, and its closure relation was proven. In a recent paper~\cite{BolPetSur2015} a more geometric interpretation of that result was provided. Here we define the general set-up for the variational approach, in terms of Lagrangian multiforms, of 3D discrete systems.

\subsection{Defining the action}
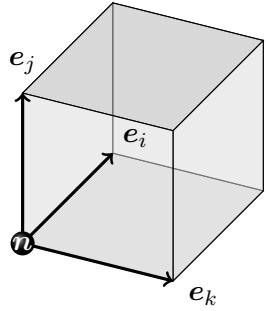
\begin{figure}[h!]\centering
\begin{tikzpicture}[every node/.style={minimum size=1cm},scale=2]
%%% back faces %%%
%% left face %%
\begin{scope}[every node/.append style={xslant=0,yslant=1},xslant=0,yslant=1]
 \filldraw[fill=gray!20,fill opacity=0.5] (0,1) rectangle (0.6,0);
\end{scope}
%% bottom face %%
\begin{scope}[every node/.append style={xslant=1,yslant=-0.2},xslant=1,yslant=-0.2]
 \filldraw[fill=gray!50,fill opacity=0.5] (0,0) rectangle (1.25,0.6);
\end{scope}
%% back face %%
 \begin{scope}[every node/.append style={xslant=0,yslant=-0.25},xslant=0,yslant=-0.25]
 \filldraw[fill=gray!20,fill opacity=0.5] (0.6,0.75) rectangle (1.6,1.75);
 %%% Note that the central point in this grid is at 0.8,0.875 %%%
\end{scope}
%%% front faces %%%
%% right face %%
\begin{scope}[every node/.append style={xslant=0,yslant=1},xslant=0,yslant=1]
 \filldraw[fill=gray!20,fill opacity=0.5] (1,-0.25) rectangle (1.6,-1.25);
\end{scope}
%% top face %%
\begin{scope}[every node/.append style={xslant=1,yslant=-0.2},xslant=1,yslant=-0.2]
 \filldraw[fill=gray!50,fill opacity=0.5] (-1,0.8) rectangle (0.25,1.4);
\end{scope}
 %% front face %%
\begin{scope}[every node/.append style={xslant=0,yslant=-0.25},xslant=0,yslant=-0.25]
 \filldraw[fill=gray!20,fill opacity=0.5] (0,0) rectangle (1,1);
\end{scope}
%% circles %%
 \draw[very thick,->] (0,0) -- (0,1);		\node at (0,1.2) {$\boldsymbol{e}_{j}$};
 \draw[very thick,->] (0,0) -- (0.6,0.6); 	\node at (0.75,0.7) {$\boldsymbol{e}_{i}$};
 \draw[very thick,->] (0,0) -- (1,-0.25);		\node at (1.2,-0.35) {$\boldsymbol{e}_{k}$};
 \drawDot{0,0}
 \node[white] at (0,0) {$\boldsymbol{n}$};
\end{tikzpicture}\vspace{-2mm}
\caption{Elementary oriented cube.}\label{nu_cube}
\end{figure}

A Lagrangian for a 3-dimensional system can be defined on an elementary cube (or rhomboid) $\nu_{ijk}(\boldsymbol{n})$, where $\nu_{ijk}(\boldsymbol{n})$ is specified by the position $\boldsymbol{n}=(\boldsymbol{n},\boldsymbol{n}+\boldsymbol{e}_{i},\boldsymbol{n}+\boldsymbol{e}_{j},\boldsymbol{n}+\boldsymbol{e}_{k})$ of one of its vertices in the lattice and the lattice directions given by the base vectors $\boldsymbol{e}_{i}$, $\boldsymbol{e}_{j}$, $\boldsymbol{e}_{k}$, as in Fig.~\ref{nu_cube}.

The Lagrangian 3-form $\mathcal{L}_{ijk}$ can depend, in principle, on the fields at all 8 vertices of the elementary cube
\begin{gather*}\label{3form}
\mathcal{L}_{ijk}(\boldsymbol{n})= L\big(u(\boldsymbol{n}),u(\boldsymbol{n}+\boldsymbol{e}_{i}),u(\boldsymbol{n}+\boldsymbol{e}_{j}),u(\boldsymbol{n}+\boldsymbol{e}_{k}),
u(\boldsymbol{n}+\boldsymbol{e}_{i}+\boldsymbol{e}_{j}), \\
\hphantom{\mathcal{L}_{ijk}(\boldsymbol{n})= L\big(}{} u(\boldsymbol{n}+\boldsymbol{e}_{j}+\boldsymbol{e}_{k}),u(\boldsymbol{n}+\boldsymbol{e}_{i}+\boldsymbol{e}_{k}),u(\boldsymbol{n}+\boldsymbol{e}_{i}+\boldsymbol{e}_{j}+\boldsymbol{e}_{k});\alpha_i,\alpha_j,\alpha_k\big),
\end{gather*}
as well as on lattice parameters $\alpha_i$, $\alpha_j$, $\alpha_k$. The action can then be defined as a connected configuration $\nu$ of these elementary cubes embedded in a lattice of 4 dimensions or higher
\begin{gather*}
 S[u(\boldsymbol{n});\nu] = \sum_{\nu_{ijk}(\boldsymbol{n})\in\nu}{\mathcal{L}_{ijk}(\boldsymbol{n})}.
\end{gather*}
This action is of course still perfectly valid if the Lagrangian doesn't depend on the fields at all vertices of the cube. For example, in the case of the bilinear discrete KP equation, one can write a Lagrangian depending on fields at 6 vertices.

\subsection{The Euler--Lagrange equations}
The Euler--Lagrange equation in the usual 3-dimensional space is
\begin{gather*}
 0 = \dfrac{\partial}{\partial u}\big(\mathcal{L}_{ijk}+T_{i}^{-1}\mathcal{L}_{ijk}+T_{j}^{-1}\mathcal{L}_{ijk}+T_{k}^{-1}\mathcal{L}_{ijk} \\
\hphantom{0=}{} +T_{i}^{-1}T_{j}^{-1}\mathcal{L}_{ijk}+T_{j}^{-1}T_{k}^{-1}\mathcal{L}_{ijk}+T_{k}^{-1}T_{i}^{-1}\mathcal{L}_{ijk}+T_{i}^{-1}T_{j}^{-1}T_{k}^{-1}\mathcal{L}_{ijk}\big),
\end{gather*}
where we take into account all Lagrangian contributions that involve the field $u$. We have suppressed the dependence on the variables, writing
\begin{gather*}
 \mathcal{L}_{ijk}=\mathcal{L}_{ijk}(u,T_{i}u,T_{j}u,T_{k}u,T_{i}T_{j}u,T_{j}T_{k}u,T_{i}T_{k}u,T_{i}T_{j}T_{k}u).
\end{gather*}
Note that any point in $\mathbb{Z}^3$ belongs to 8 cubes, so we have in principle 8 terms in the above equation. This is the analogue of the ``flat'' equation~\eqref{EL2d} we had in 2 dimensions.

Embed the system in 4 dimensions. In 3 dimensions, the smallest closed 2-dimensional space is a cube, consisting of 6 faces; in 4 dimensions, the smallest closed 3-dimensional space is a~hypercube, consisting of 8 cubes. The action on the elementary hypercube will have the form
\begin{gather*}
 S(u;{\rm hypercube}) = \Delta_{l}\mathcal{L}_{ijk}-\Delta_{i}\mathcal{L}_{jkl}+\Delta_{j}\mathcal{L}_{kli}-\Delta_{k}\mathcal{L}_{lij}.
\end{gather*}
Because of the symmetry, we need only take derivatives with respect to $u,T_{i}u$, $T_{i}T_{j}u$, $T_{i}T_{j}T_{k}u$ and $T_{i}T_{j}T_{k}T_{l}u$, and the other equations will follow by cyclic permutation of the lattice directions. Then we have the set of equations
\begin{subequations}\label{3deqn}
\begin{gather}
 0 = \frac{\partial}{\partial u}\big({-}\mathcal{L}_{ijk}+\mathcal{L}_{jkl}-\mathcal{L}_{kli}+\mathcal{L}_{lij}\big),\label{3deqna}\\
 0 = \frac{\partial}{\partial T_{i}u}\big({-}\mathcal{L}_{ijk}-T_{i}\mathcal{L}_{jkl}-\mathcal{L}_{kli}+\mathcal{L}_{lij}\big),\label{3deqnb}\\
 0 = \frac{\partial}{\partial T_{i}T_{j}u}\big({-}\mathcal{L}_{ijk}-T_{i}\mathcal{L}_{jkl}+T_{j}\mathcal{L}_{kli}+\mathcal{L}_{lij}\big),\label{3deqnc}\\
 0 = \frac{\partial}{\partial T_{i}T_{j}T_{k}u}\big({-}\mathcal{L}_{ijk}-T_{i}\mathcal{L}_{jkl}+T_{j}\mathcal{L}_{kli}-T_{k}\mathcal{L}_{lij}\big),\label{3deqnd}\\
 0 = \frac{\partial}{\partial T_{i}T_{j}T_{k}T_{l}u}\big(T_{l}\mathcal{L}_{ijk}-T_{i}\mathcal{L}_{jkl}+T_{j}\mathcal{L}_{kli}-T_{k}\mathcal{L}_{lij}\big),\label{3deqne}
\end{gather}
\end{subequations}
along with the equivalent shifted versions
\begin{subequations}\label{3deqnshift}
\begin{gather}
 0 = \frac{\partial}{\partial u}\big({-}\mathcal{L}_{ijk}+\mathcal{L}_{jkl}-\mathcal{L}_{kli}+\mathcal{L}_{lij}\big),\\
 0 = \frac{\partial}{\partial u}\big({-}T_{i}^{-1}\mathcal{L}_{ijk}-\mathcal{L}_{jkl}-T_{i}^{-1}\mathcal{L}_{kli}+T_{i}^{-1}\mathcal{L}_{lij}\big),\\
 0 = \frac{\partial}{\partial u}\big({-}T_{i}^{-1}T_{j}^{-1}\mathcal{L}_{ijk}-T_{j}^{-1}\mathcal{L}_{jkl}+T_{i}^{-1}\mathcal{L}_{kli}+T_{i}^{-1}T_{j}^{-1}\mathcal{L}_{lij}\big),\\
 0 = \frac{\partial}{\partial u}\big({-}T_{i}^{-1}T_{j}^{-1}T_{k}^{-1}\mathcal{L}_{ijk}-T_{j}^{-1}T_{k}^{-1}\mathcal{L}_{jkl}+T_{i}^{-1}T_{k}^{-1}\mathcal{L}_{kli}-T_{i}^{-1}T_{j}^{-1}\mathcal{L}_{lij}\big),\\
 0 = \frac{\partial}{\partial u}\big(T_{i}^{-1}T_{j}^{-1}T_{k}^{-1}\mathcal{L}_{ijk}-T_{j}^{-1}T_{k}^{-1}T_{l}^{-1}\mathcal{L}_{jkl}+T_{i}^{-1}T_{k}^{-1}T_{l}^{-1}\mathcal{L}_{kli}-T_{i}^{-1}T_{j}^{-1}T_{l}^{-1}\mathcal{L}_{lij}\big).
\end{gather}
\end{subequations}
The system of equations comprising \eqref{3deqn}, \eqref{3deqnshift} and the 3-form closure relation
\begin{gather}\label{3dclosure}
 \Delta_{l}\mathcal{L}_{ijk}-\Delta_{i}\mathcal{L}_{jkl}+\Delta_{j}\mathcal{L}_{kli}-\Delta_{k}\mathcal{L}_{lij}=0
\end{gather}
forms the fundamental system of EL equations of a 3D integrable fully discrete system, which we leave without proof for now. The, so far only, example that has been realized is the case of the bilinear discrete KP, which we will present next.

\subsection{Example: bilinear discrete KP}
The Lagrangian for the bilinear discrete KP equation was first given in \cite{LobNijQui2009}, and in 3-dimensional space gives as Euler--Lagrange equations 12 copies of the bilinear discrete KP equation itself, on 6 elementary cubes. The Lagrangian $\mathcal{L}_{ijk}$ 3-form component depends on the six fields $T_{i}u$, $T_{j}u$, $T_{k}u$, $T_{i}T_{j}u$, $T_{j}T_{k}u,$ and $T_{i}T_{k}u$, and has the following form
\begin{subequations}\label{L123}\begin{gather}
 \mathcal{L}_{ijk} = \frac{1}{2}\bigl(L_{ijk}+L_{jki}+L_{kij}-L_{ikj}-L_{jik}-L_{kji}\bigr), \label{L123a}
 \end{gather}
 where
\begin{gather} L_{ijk}:= \ln\left(\frac{(T_{k}u)T_{i}T_{j}u}
 {(T_{j}u)T_{k}T_{i}u}\right)\ln\left(-\frac{A_{ki}T_{j}u}{A_{jk}T_{i}u}\right)
 -{\rm Li}_{2}\left(-\frac{A_{ij}(T_{k}u)T_{i}T_{j}u}{A_{ki}(T_{j}u)T_{k}T_{i}u}\right). \label{L123b}
\end{gather}\end{subequations}
Here the $A_{ij}$ are constants which are antisymmetric with respect to swapping the indices, i.e., $A_{ij}=-A_{ji}$. Introducing the quantity $C_{ijk}$, defined by
\begin{gather*}
 C_{ijk} = \frac{A_{ij}(T_{k}u)T_{i}T_{j}u+A_{jk}(T_{i}u)T_{j}T_{k}u}{A_{ik}(T_{j}u)T_{k}T_{i}u},
\end{gather*}
the single elementary block $L_{ijk}$ viewed as a Lagrangian in its own right would yield the following Euler--Lagrange equation
\begin{gather*}
 \frac{1}{u}\ln\left\{\frac{\big(T_{k}^{-1}C_{kij}\big)T_{i}^{-1}T_{j}^{-1}C_{kij}}{\big(T_{j}^{-1}C_{kij}\big)T_{k}^{-1}T_{i}^{-1}C_{kij}}\right\}=0,
\end{gather*}
which comprises four copies of the Hirota's bilinear $\Delta$KP equation, given by $C_{kij}=1$.

\newcommand{\drawLinewithGBG}[2]
{
 \draw[gray!20,line width=3pt] (#1) -- (#2);
 \draw[black,very thick] (#1) -- (#2);
}
\newcommand{\drawLinewithNBG}[2]
{
 \draw[black,very thick] (#1) -- (#2);
}
\newcommand{\drawArrow}[2]
{
 \draw[black,very thick,->] (#1) -- (#2);
}

\begin{figure}[h]\centering
%%%%%%%%%%%%%%%%%%%
%%% BIG PICTURE %%%
%%%%%%%%%%%%%%%%%%%
\begin{tikzpicture}[every node/.style={minimum size=1cm},scale=2,>=stealth]
%%% Arrows %%%
 \drawArrow{-1.2,0.475}{-0.7,0.375}
 \drawArrow{-1.2,0.475}{-0.9,0.775}
 \drawArrow{-1.2,0.475}{-1.2,0.975}
 \node at (-0.6,0.35) {$n_{i}$};
 \node at (-0.8,0.8) {$n_{j}$};
 \node at (-1.2,1.05) {$n_{k}$};
%%% back faces %%%
%% left face %%
\begin{scope}[every node/.append style={xslant=0,yslant=1},xslant=0,yslant=1]
 \drawGridLeft
\end{scope}
%% bottom face %%
\begin{scope}[every node/.append style={xslant=1,yslant=-0.2},xslant=1,yslant=-0.2]
 \drawGridBottom
\end{scope}
 %% back face %%
 \begin{scope}[every node/.append style={xslant=0,yslant=-0.25},xslant=0,yslant=-0.25]
 \drawGridBack
 \drawLinewithGBG{0.3,0.875}{0.3,1.375}
 \drawLinewithGBG{0.3,0.875}{0.6,1.25}
 \drawLinewithGBG{0.5,0.5}{0.8,0.875}
 \drawLinewithGBG{0.6,0.75}{1.1,0.75}
 \drawLinewithGBG{0.6,1.25}{0.6,0.75}
 \drawLinewithGBG{0.6,1.25}{0.6,1.75}
 \drawLinewithGBG{0.8,0.375}{0.8,0.875}
 \drawLinewithGBG{0.8,0.375}{1.3,0.375}
 \drawLinewithGBG{0.8,0.875}{0.3,0.875}
 \drawLinewithGBG{0.8,0.875}{1.1,1.25}
 \drawLinewithGBG{1.1,0.75}{0.8,0.375}
 \drawLinewithGBG{1.1,0.75}{1.6,0.75}
 \drawLinewithGBG{1.1,1.25}{0.6,1.25}
 \drawLinewithGBG{1.1,1.25}{1.1,0.75}
 \drawLinewithGBG{1.3,0.875}{0.8,0.875}
 \drawLinewithGBG{0.8,1.375}{0.8,0.875}
\end{scope}
%% circles %%
 \drawDot{1.1,0.975}
 \drawDot{1.1,0.475}
 \drawDot{0.8,0.175}
 \drawDot{0.6,1.1}
 \drawDot{0.6,0.6}
 \drawDot{0.3,0.8}
%% back face (that sneaky line that should be in front of one circle) %%
\begin{scope}[every node/.append style={xslant=0,yslant=-0.25},xslant=0,yslant=-0.25]
 \drawLinewithGBG{0.8,0.875}{0.5,0.5}
\end{scope}
\drawDiamond{0.8,0.675}
%%% front faces %%%
%% right face %%
\begin{scope}[every node/.append style={xslant=0,yslant=1},xslant=0,yslant=1]
 \drawGridRight
\end{scope}
%% top face %%
\begin{scope}[every node/.append style={xslant=1,yslant=-0.2},xslant=1,yslant=-0.2]
 \drawGridTop
\end{scope}
 %% front face %%
\begin{scope}[every node/.append style={xslant=0,yslant=-0.25},xslant=0,yslant=-0.25]
 \drawGridFront
 \drawLinewithNBG{1.6,0.75}{1.3,0.375}
 \drawLinewithGBG{1.3,0.375}{1.3,0.875}
 \drawLinewithNBG{0.3,1.375}{0,1}
 \drawLinewithGBG{0,1}{0.5,1}
 \drawLinewithGBG{0.5,1}{0.5,0.5}
 \drawLinewithGBG{0.8,1.375}{0.5,1}
 \drawLinewithGBG{0.5,1}{1,1}
 \drawLinewithGBG{1,1}{1,0.5}
 \drawLinewithGBG{1,0.5}{1.3,0.875}
 \drawLinewithNBG{0.6,1.75}{0.3,1.375}
 \drawLinewithGBG{0.3,1.375}{0.8,1.375}
 \drawLinewithNBG{1.3,0.375}{1,0}
 \drawLinewithGBG{1,0}{1,0.5}
 \drawLinewithGBG{1,0.5}{0.5,0.5}
\end{scope}
%% circles %%
 \drawDot{1.6,0.375}
 \drawDot{1.3,0.575}
 \drawDot{1.3,0.075}
 \drawDot{1,0.775}
 \drawDot{1,0.275}
 \drawDot{1,-0.225}
 \drawDot{0.8,1.175}
 \drawDot{0.6,1.6}
 \drawDot{0.5,0.875}
 \drawDot{0.5,0.375}
 \drawDot{0,1}
 \drawDot{0.3,1.3}
\end{tikzpicture}\vspace{-2mm}
\caption{The 19-point equation $\delta \mathcal{L}_{ijk}/\delta\tau=0$.}\label{bigpicture}
\end{figure}
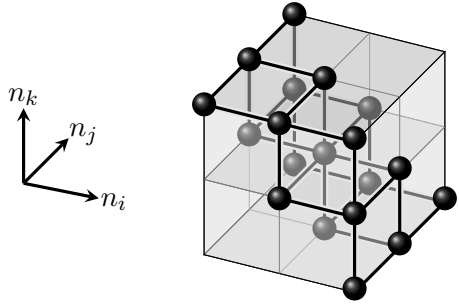

As a consequence the fully antisymmetric Lagrangian 3-form component $\mathcal{L}_{ijk}$ of \eqref{L123}, considered as a scalar Lagrangian defined in $\mathbb{Z}^3$ lattice, by variation of the corresponding lattice action yields an equation combining 12 shifted copies of the $\Delta$KP equation, namely
\begin{gather*}
 0 = \frac{1}{u}\ln\left\{\frac{\big(T_{k}^{-1}C_{kij}\big)T_{i}^{-1}T_{j}^{-1}C_{kij}}{\big(T_{j}^{-1}C_{kij}\big)T_{k}^{-1}T_{i}^{-1}C_{kij}}\cdot
 				 \frac{\big(T_{i}^{-1}C_{ijk}\big)T_{j}^{-1}T_{k}^{-1}C_{ijk}}{\big(T_{k}^{-1}C_{ijk}\big)T_{i}^{-1}T_{j}^{-1}C_{ijk}}\cdot
				 \frac{\big(T_{j}^{-1}C_{jki}\big)T_{k}^{-1}T_{i}^{-1}C_{jki}}{\big(T_{i}^{-1}C_{jki}\big)T_{j}^{-1}T_{k}^{-1}C_{jki}}\right\},
\end{gather*}
 which forms a 19-point equation involving the configuration of points illustrated in Fig.~\ref{bigpicture}. (See~\cite{LobNijQui2009} for figures illustrating how the twelve 6-point equations of the bilinear discrete KP equation are embedded in this configuration.)

Let us now consider the system of 3-form Euler--Lagrange equations~\eqref{3deqn} and~\eqref{3deqnshift}. Equations~\eqref{3deqna} and \eqref{3deqne} are trivial in the case of the Lagrangian 3-form given by~\eqref{L123}, while \eqref{3deqnb}--\eqref{3deqnd} yield
\begin{gather*}
 0 = \frac{1}{T_{i}u}\ln\left\{\frac{C_{ijk}C_{jli}C_{ikl}}{C_{jki}C_{ijl}C_{kli}}\right\},\\
 0 = \frac{1}{T_{i}T_{j}u}\ln\left\{\frac{C_{kij}C_{ijl}T_{i}C_{jkl}T_{j}C_{kli}}{C_{ijk}C_{lij}T_{i}C_{klj}T_{j}C_{ikl}}\right\},\\
 0 = \frac{1}{T_{i}T_{j}T_{k}u}\ln\left\{\frac{T_{i}C_{ljk}T_{k}C_{lij}T_{j}C_{ikl}}{T_{i}C_{jkl}T_{k}C_{ijl}T_{j}C_{lik}}\right\}.
\end{gather*}
while the closure relation \eqref{3dclosure} was proven in \cite{LobNijQui2009}. A geometrical interpretation of the Lagrangian structure of the discrete KP equation was given in the recent paper~\cite{BolPetSur2015}.

\section{Summary and conclusions}\label{section_conclusion}
Multidimensionally consistent systems can be considered as critical points of an action: critical with respect to the dependent variable, and also with respect to the curve or surface in the space of independent variables. In the case of discrete systems, this means the action is required to be independent of the curve or surface on which it is defined, whilst keeping any boundary it may have fixed. This leads to a set of Euler--Lagrange equations, corresponding to basic configurations of points in a surface, which should be satisfied simultaneously.

In the case of 2-dimensional discrete systems defined on quad-lattices, we have shown that the extended system of Euler--Lagrange equations arising from this variational principle specify a general form of the Lagrangian (up to some essential freedom, which is allowed within the multidimensional consistency) and how quadrilateral lattice equations arise by a procedure analogous to integration by quadratures.

As we pointed out, we consider the set of EL equations as a defining system of equations for the Lagrangian itself. This constitutes a significant departure from the conventional point of view where the Lagrangian is given (usually obtained from considerations from physics, symmetry or other tertiary considerations), and thus marks in our view a new vision on how we should regard the variational imperative: in the integrable case of Lagrangian multiforms, the Lagrangians themselves are part of the solution of the extended system of equations obtained from varying not only the field variables on a given space-time of independent variables, but by also varying the geometry of space-time itself. It would be of interest to see whether Lagrangians associated with descriptions of known physical processes could be obtained from such a~novel variational theory. In this respect the most important direction to pursue is the setting up of a quantum theory associated with the Lagrangian multiform theory. A first step in this direction for the case of quadratic Lagrangians was done in the recent paper~\cite{KingNij17}.

It would also be interesting to see if the results of this paper can be extended to higher than 3 dimensions where we will have a~Lagrangian function evaluated on an $n$-dimensional object, in particular on an $n$-dimensional cube. Embedding this in higher dimensions, we consider an action on the smallest closed $n$-dimensional surface in $(n+1)$ dimensions, a hypercube. Then the minimal set of Euler--Lagrange equations are obtained by demanding that the derivative of this action with respect to each variable is zero.

\subsection*{Acknowledgements}
The authors would like to thank James Atkinson for helpful comments and suggestions. At the time of writing SL was supported by the Australian Laureate Fellowship Grant \#FL120100094 from the Australian Research Council. FN was partially supported by the grant EP/I038683/1 of the Engineering and Physical Sciences Research Council (EPSRC). FN is also grateful to the hospitality of the Sophus Lie Center in Nordfjordeid (Norway) during the conference on ``Nonlinear Mathematical Physics: Twenty Years of JNMP'' (June 4--14, 2013) where a~pre\-liminary account (joint with SL) of the results of this paper was first presented \cite{Nij2013} prior to the appearance of other papers reporting similar results.

\pdfbookmark[1]{References}{ref}
 \LastPageEnding

\end{document}